\documentclass[11pt,twoside]{article}

\usepackage{asp2006}
\usepackage{epsf}
\usepackage{psfig}
\usepackage{lscape}

\def\jcoph{J. Comp.\ Phys.}

\def\eg{{\it e.g.}}
\def\etal{{\it et al.}}
\def\etc{{\it etc.}}

\def\pmb#1{\setbox0=\hbox{$#1$}%
  \kern-0.25em\copy0\kern-\wd0
  \kern.05em\copy0\kern-\wd0
  \kern-0.025em\raise.0433em\box0}
\long\def\Ignore#1{\relax}

\begin{document}

\title{Dynamical Evolution of Disk Galaxies}
\author{J. A. Sellwood}
\affil{Rutgers University, Department of Physics \& Astronomy, \\
       136 Frelinghuysen Road, Piscataway, NJ 08854-8019 \\
       {\it sellwood@physics.rutgers.edu}}

\bigskip
\centerline{\it Rutgers Astrophysics Preprint \#479}

\begin{abstract}
Spiral patterns are important agents of galaxy evolution.  In this
review, I describe how the redistribution of angular momentum by
recurrent transient spiral patterns causes the random speeds of stars
to rise over time, metallicity gradients to be reduced, and drives
large-scale turbulence in the disk, which could be important for
galactic dynamos.  I also outline a possible mechanism for the
recurrence of spiral instabilities and supporting evidence from solar
neighborhood kinematics.  Finally, I confirm that cloud scattering
alone would predict the local velocity ellipsoid to be flattened,
contrary to long-held expectations.
\end{abstract}

\section{Introduction}
\label{intro}
The present-day properties of disk galaxies are a consequence both of
the initial conditions under which they formed and their dynamical
evolution, which is affected by both internally- and externally-driven
processes.  In this review, I focus principally on the evolution of an
isolated disk, especially the dynamical effects of spiral patterns.

Historical computer simulations of isolated, globally-stable,
self-gravitating disks (Miller, Prendergast \& Quirk 1970; Hockney \&
Brownrigg 1974; \etc)\ exhibited transient spiral patterns for at
least a few tens of galaxy rotations.  Modern simulations of forming
galaxies (\eg, Gottl\"ober \etal\ 2002; Abadi \etal\ 2003; Governato
\etal\ 2004, 2007) exhibit similar behavior, but include so many
different physical processes that it is unclear which is responsible
for the spiral patterns.

It has long been hoped that the spontaneous formation of short-lived
spiral patterns in simulations reflects the behavior in real galaxies,
although our single snapshot view of every galaxy precludes a direct
determination of the lifetimes of individual spiral patterns in
galaxies (but see Meidt \etal\ 2007).  An alternative theory (\eg\
Bertin \& Lin 1996) argues that most spirals should be long-lived,
quasi-stationary features, but no observational test has yet been able
to determine which picture is correct.  The difference between these
two points of view is important since the predictions for the
evolution of galaxy disks differ substantially.

\section{Dynamical consequences of short-lived spiral patterns}
Angular momentum transport is the principal driver of galaxy evolution
by spiral patterns.  Not only does it change the radial distribution
of mass within the disk, especially near the outer edge (\eg\ Ro\u
skar \etal\ 2008), but it also drives increasing velocity dispersions
and radial mixing within the disk.  In all these ways, recurrent
transient spiral patterns have a greater impact on the evolution of
galaxy disks than do long-lived waves.

It might seem that if the time-averaged amplitude and pitch angle of
transient spirals did not differ much from those of a steady
long-lived pattern, the gravity torques and angular momentum changes
would be similar.  However, the gravitational stresses caused by
long-lived patterns that invoke feed-back via the long-wave branch of
the dispersion relation (see ch. 6 of Binney \& Tremaine 2008,
hereafter BTII) transport less angular momentum because the
complicating advective transport term (aka ``lorry transport''
Lynden-Bell \& Kalnajs 1972; BTII, Appendix J) greatly reduces the
angular momentum flux.  Furthermore, since evolutionary angular
momentum exchanges between the stars and the wave take place at
resonances, large changes induced by a long-lived pattern would be
confined to a few narrow resonances, the most important of which,
the inner Lindblad resonance, is ``shielded.''  On the other hand,
multiple, short-lived disturbances having a range of pattern speeds
over the same time interval share the exchanges over large parts of
the disk because their resonances are broader and have more numerous
locations.

\subsection{The importance of gas}
Sellwood \& Carlberg (1984) showed that recurrent transient spiral
patterns fade after about 10 galaxy rotation periods because the
time-dependent gravitational potential fluctuations caused by the
spirals themselves scatter the disk particles away from circular
orbits at broad resonances (Carlberg \& Sellwood 1985).  It becomes
harder to organize coherent spiral density variations as the velocity
dispersion of the disk particles rises; spiral activity in a disk of
particles is therefore self-limiting.  They also showed that spirals
could recur indefinitely only if some of the effects of gas
dissipation and star formation were included in the calculations (see
also Carlberg \& Freedman 1985 and Toomre 1990), and the results from
modern work are consistent with this picture.  The build up of random
motion is resisted in the dissipative component, and the formation of
new stars with small peculiar velocities at the rate of a few per year
is enough to maintain spiral activity in the entire disk.

Spiral patterns are prominent only in galaxies containing significant
gas from which stars are forming; lenticular galaxies are mostly gas
free and have little in the way of spiral features.  Recurrent
transient spiral behavior offers a natural, though not the only
possible, explanation of this fact.

\subsection{Age-velocity dispersion relation}
The velocity dispersions of solar neighborhood stars reveal an
increasing trend when they are arranged by age (Wielen 1977;
Nordstr\"om \etal\ 2004) or, for main-sequence stars, by color (Dehnen
\& Binney 1998).  I am unqualified to contribute to the on-going
discussion (\eg\ Holmberg \etal\ 2007 and references therein) of the
accuracy of stellar age estimates adopted by Edvaardson \etal\ (1993)
and Nordstr\"om \etal\ (2004), but errors of order unity would be
required to totally vitiate the dynamical significance of their
results.  Possible smaller errors of up to 20\% in individual stellar
ages do not affect the dynamical implications of their results.

Spitzer \& Schwarzschild (1953) proposed scattering by massive clouds
of gas, before even their discovery, as the dynamical origin of the
general increase of random motion with age -- the age-velocity
dispersion relation.  Lacey (1984) extended their analysis to 3-D and
concluded (see also Lacey 1991) that the observed molecular clouds
could not account for the magnitude of the increase to that of the
oldest stars.

Stars are also scattered by transient spiral waves (Barbanis \&
Woltjer 1967; Carlberg \& Sellwood 1986).  Lacey (1991) concluded that
transient spirals were a promising mechanism to account for the higher
dispersion of the older stars, although other ideas may not be ruled
out.

\subsection{Dispersion of stellar metallicities with age}
Edvaardson \etal\ (1993), and others, have reported that older stars
in the solar neighborhood have a spread of metallicities that is
inconsistent with the idea that they all formed at the solar radius
from gas that was gradually becoming more metal rich over time.  This
result is more critically dependent on age estimates, although errors
would have to be of order unity to reduce the reported spread to a
tight correlation, which would also make the Sun a truly exceptional
star.

Sellwood \& Binney (2002) showed that radial migration of stars driven
by recurrent, transient spiral waves in fact provides a natural
explanation for the metallicity spread of stars with age.  They showed
that the galactocentric radius of a star can migrate by up to 2~kpc in
either direction as a result of ``surfing'' near the corotation radius
of an individual spiral pattern.  The combined effects of multiple
spiral patterns leads to radial mixing of stars without an associated
increase in the velocity dispersion.  Long-lived spiral waves would
not achieve a quasi-steady diffusion, since stars on horse-shoe type
orbits near corotation would alternate between two mean radii,
preventing the radial diffusion that is needed to account for the
increasing range of metallicities with age.

\subsection{Large-scale turbulence}
Precisely the same mechanism that causes radial mixing of the stars
creates large-scale turbulence in the ISM.  Gas in the vicinity of
corotation is driven by a spiral pattern, radially inwards by $\sim
1\;$kpc at some azimuths and outwards by a similar amount at others.
The radially-shifted gas eventually mixes with other gas (\eg\
Sellwood \& Preto 2002, fig.\ 9) at its new radius.

Spiral-driven mixing in the ISM may also help with the well-known
problem posed by the large-scale (ordered) component of {\bf B}-fields
in galaxies (\eg\ Rees 1994).  Standard $\alpha\Omega$-dynamo theory
(Parker 1955) is thought to yield too low a growth rate to achieve the
present-day observed field strengths (Beck \etal\ 1996) from the
likely seed fields.  The growth-rate is proportional to the geometric
mean of the rates of galactic shear (the $\Omega$ term) and cyclonic
circulation (the $\alpha$ term) (Kulsrud 1999).  Current estimates of
the $\alpha$-term are based on supernovae-driven turbulence
(Ferri\`ere 1998; Balsara \& Kim 2005), but spiral-driven turbulence
should enhance the $\alpha$-effect substantially, and thereby increase
the growth-rate obtainable from the dynamo.

\section{Mechanism for recurrent transient spirals}
The spiral patterns in computer simulations of isolated disks must be
self-excited.  However, without an understanding of their origin, they
could all be dismissed as arising from some unknown numerical
artifact, perhaps related to the small number of particles (Bertin \&
Lin 1996).  If true, such a skeptical viewpoint would require spirals
in real galaxies to have a different origin, though they need not
necessarily be long-lived; a possible alternative universal mechanism
for recurrent transient spirals could be forcing by substructure in
the dark matter halos, as reported by Dubinski (these proceedings).

The development of short-lived spiral features has not changed over
the years as the numerical quality of simulations has improved.  Codes
have held up under extensive testing (Sellwood 1983; Inagaki, Nishida
\& Sellwood 1984; Sellwood \& Athanassoula 1986; Earn \& Sellwood
1995; Sellwood \& Evans 2001), but a successful test in one problem is
no guarantee of the code's performance in other problems.  I have
recently (Sellwood, in preparation) conducted a suite of simulations
of a stable Mestel disk model with particle numbers ranging up to $N=5
\times 10^8$.  Simulations of a linearly stable disk should manifest
no structure exceeding that expected from amplified particle noise.
However, I find evidence that the linear theory prediction fails
whenever the disturbance amplitude exceeds $\sim 2$\% of the
undisturbed density.  Again this could be an important physical result
or another manifestation of the supposed artifact, although the
greatest particle number is now only 2 orders of magnitude below the
number of stars in a disk.

The most effective way to counter this dismissive viewpoint would be
to understand the mechanism for recurrent spiral generation in
simulations and to find evidence that the process also works in
nature.  I (Sellwood 2000) outlined a possible mechanism for recurrent
spiral generation in collisionless particle disks, but observational
evidence to support it was lacking.  However, the detailed phase-space
structure of stars in the solar neighborhood, as revealed in the
monumental study of local F \& G dwarfs by Nordstr\"om \etal\ (2004)
has opened the door to empirical tests.  Clearly, data from the
upcoming GAIA mission will be even more useful.

The central idea of the recurrence mechanism is that scattering of
stars at the principal resonances of the spiral pattern (\eg\ BTII)
changes the distribution function (DF) in such a way as to seed the
growth of a new instability in the disk.  Each instability is caused
by locally steep gradients in the DF, supported vigorously by the
response of the surrounding disk (Sellwood \& Kahn 1991).  The spiral
wave grows until non-linear effects become important, and Sellwood \&
Binney (2002) showed that the onset of horse-shoe orbits at corotation
causes the disturbance to begin to disperse.  At this moment, all the
action stored in the wave is carried away to the Lindblad resonances
at the group velocity where stars are scattered to create the
conditions for a new instability.  While the idea is simply stated,
many of the details remain to be worked out.

Support for this idea comes from analysis of the velocity distribution
of $13\,240$ nearby F \& G dwarf stars in the Geneva-Copenhagen sample
(Nordstr\"om \etal\ 2004).  The distribution of stars in phase space
is very far from smooth, as originally noted by Dehnen (1998), but is
broken into various ``star streams,'' which cannot simply be dissolved
star clusters (Famaey \etal\ 2006; Bensby \etal\ 2007), with no
underlying smooth component.

Some of these features have been modeled as scattering by the bar
(Kalnajs 1991; Dehnen 2000) or by spirals (Yu \& Tremaine 2002) or a
combination of both (Quillen \& Minchev 2005), but all these studies
modeled the velocity-space distribution, not integral-space, and none
was self-consistent.  The substructure in the DF is too complicated to
have a single origin, and such ideas could be responsible for one or
more of the features.

I find (in preparation) that the local DF, when plotted as a function
of energy and angular momentum, contains at least one clear feature of
high statistical significance that appears to have resulted from
either scattering or trapping at a resonance.  Analysis of the sample
in action-angle variables reveals that it is an inner Lindblad
resonance that has most recently sculptured the local DF, which is
exactly the feature Sellwood (1994) predicted might be observable in
such a sample.

If the mechanism invoked to interpret the behavior of simulations does
indeed appear to occur in the Milky Way, then the entire picture stands
on much firmer ground.

\section{Velocity Ellipsoid Shape}
I conclude with an attempt to resolve an old dispute about the axis
ratio of the local velocity ellipsoid expected from scattering by
giant molecular clouds.  Though a side issue to the main thread of
this brief review, it has provided an additional, apparently
compelling, argument in favor of spirals as short-lived transients
that now turns out to have no weight.

Lacey (1984) and Binney \& Lacey (1988) calculated the separate growth
of the in-plane and vertical components, concluding that cloud
scattering should cause the vertical component to be intermediate
between the radial and azimuthal components.  Their result seems
physically plausible on energy equipartition grounds: scattering by
massive clouds redirects the peculiar motions of stars through random
angles, and therefore isotropizes the motions as far as is possible.

However, this is not what is observed.  The second moments of the
velocity distribution of solar neighborhood stars in the three
orthogonal directions are unequal (Wielen 1977; Dehnen \& Binney 1998;
Nordstr\"om \etal\ 2004).  The radial component, $\sigma_u$, is the
largest, the azimuthal component, $\sigma_v$, intermediate and the
vertical component, $\sigma_w$, is the smallest, and this remains true
for all groups when the stars are subdivided according to the best
estimates of their ages.  The ratio of the two in-plane components is
set by the epicyclic motions, and is in reasonable agreement with
theoretical expectations (\eg\ Dehnen \& Binney 1998).  Some estimates
of the velocity ellipsoid shape in external galaxies are beginning to
be made (Gerssen \etal\ 1997, 2000; Ciardullo \etal\ 2004; Westfall
\etal, these proceedings).

Carlberg (1987) and Jenkins \& Binney (1990) therefore developed the
plausible argument that spirals drive up the in-plane components more
rapidly than scattering is able to redirect those motions into the
vertical direction, thereby accounting for the observed axis ratios of
the velocity ellipsoid.  Their argument appeared to offer strong
support for the transient spiral picture, but this particular line of
argument now seems to be incorrect.

The conclusion by Lacey (1984) and Binney \& Lacey (1988) that cloud
scattering would lead to the vertical dispersion being intermediate
between the radial and azimuthal dispersions was challenged by Ida,
Kokuba \& Makino (1993).  These authors claimed that cloud scattering
alone would lead to the vertical component being the smallest, as
confirmed in numerical simulations (Shiidsuke \& Ida 1999).  Ida
\etal\ ( 1993) also show how the axis ratio depends on the local slope
of the rotation curve.  Their two papers have attracted little
attention, largely because they are hard to follow, and their physical
explanation for their different conclusion is rather enigmatic.
Accordingly, I here describe some idealized simulations to determine
what should be the shape of the velocity ellipsoid due to scattering
by randomly distributed, co-orbiting mass clumps.

\subsection{Idealized simulations}
I wish to study the motion of collisionless stars in an axisymmetric
disk potential that is smooth except for the presence of a collection
of non-interacting scattering masses.  I treat the stars as test
particles moving in an axisymmetric, vertically stratified, potential
and integrate their motion subject to perturbations from a collection
of co-moving heavy particles.  Following Wisdom \& Tremaine (1988) for
a Keplerian disk and Toomre \& Kalnajs (1991) for a constant-velocity
disk, I consider a small orbiting patch of the disk, with periodic
boundary conditions and adjacent radial boxes sliding past the main
box at the local shear rate.  These authors describe the set-up of
sheared sheet simulations more fully.

I assume a flat rotation curve for the in-plane accelerations, while
the vertical density profile has the form $\rho(z) = \rho_0 \; {\rm
sech}^2(z/2z_0)$ for the isothermal sheet (Spitzer 1942; BTII).  The
heavy particles, which contain 20\% of the local surface density,
affect the vertical balance and I reduce the nominal surface mass
density that gives rise to the fixed potential by the mass in the
heavies, which are distributed in a thinner layer.  The heavy
particles all start with random motions that are 5\% of those of the
test particles, with a corresponding reduction in the thickness of the
layer of heavies.  I have verified that the final shape of the
velocity ellipsoid is insensitive to the fraction of mass in the heavy
particles or their velocity dispersions.

The initial in-plane velocities of the test particles are set such
that, were the disk self-gravitating, Toomre's $Q=1$ and the epicycle
approximation requires $\sigma_u = 2^{1/2}\sigma_v$.  There can be
no asymmetric drift in the sheared sheet, since the symmetry of the
equations does not determine the direction to the center of the
galaxy.  The initial vertical positions and speeds of all test
particles are set to ensure vertical balance.  The principal parameter
I vary is the initial vertical velocity dispersion of the test
particles.  The scale height is related to the dispersion as $z_0 =
\sigma_w^2 / (2\pi G\Sigma)$.  As the vertical motions of the test
particles rise, I make corresponding adjustments to the scale height
of the potential, although the final shape of the velocity ellipsoid
is little affected by turning this refinement on or off.

I adopt the usual Plummer potential for each heavy particle, with a
softening length of $0.01R_0$, where $R_0$ is our fourth the radial
extent of the sheared patch under study.  The maximum range of the
perturbing forces from the heavy particles in the adopted shearing box
arrangement that can be conveniently handled is the radial extent of
the box.  The position of every heavy particle is duplicated in the
surrounding boxes and the perturbing forces from all the heavy
particles and their images separately affect the motion of the test
particles in the main box.

\subsection{Results}
As the integration proceeds, the shape of the velocity ellipsoid of
the test particles evolves towards a steady value, while the total
magnitude of the velocity dispersion rises slowly.  I find that
irrespective of whether I begin with a round or strongly flattened
velocity ellipsoid, it quickly evolves to a shape in which the
vertical dispersion is slightly smaller than the azimuthal dispersion.
The final ellipsiod shape is always $\sigma_u : \sigma_v : \sigma_w
\simeq 1 : 0.71 : 0.62$, confirming the prediction by Ida and
coworkers, and in disagreement with the results by Lacey and Binney \&
Lacey.  The physically plausible energy equipartition argument must
also be misleading.

\begin{figure}[t]
\centerline{\psfig{file=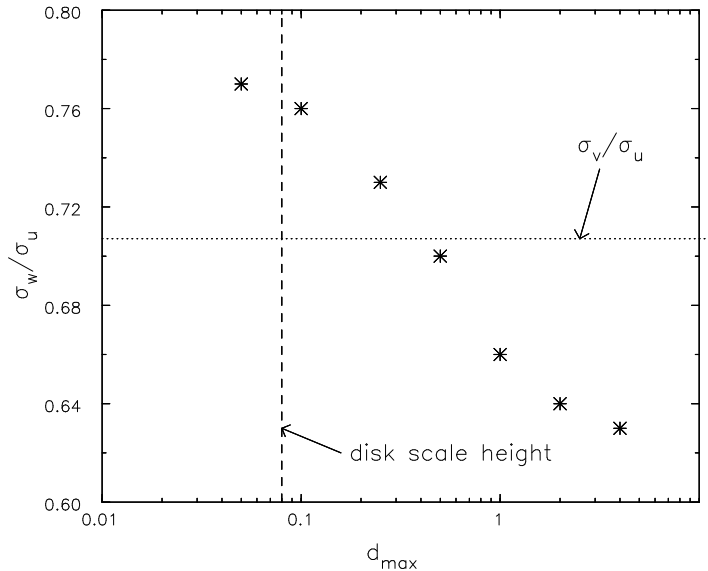,width=.7\hsize,clip=}}
\caption{\footnotesize The effect on the equilibrium axis ratio of the
velocity ellipsoid of limiting the range of the perturbation forces
from the heavy particles.}
\label{vrange}
\end{figure}

It seemed possible that this prediction fails in a thin disk because
the distribution of scatterers is not uniform.  As with any scattering
process in 3-D, the $\ln\Lambda$ term implies that distant encounters
dominate.  But distant scatterers in the flattened geometry of a disk
are not isotropically distributed, and scattering by distant clouds
must predominantly affect the in-plane star velocities, and couple
much less strongly to the vertical component.

In order to test this idea, I experimented with excluding the
influence of the disturbance forces from all heavy particles more
distant than $d_{\rm max}$.  Figure~\ref{vrange} shows the equilibrium
ratio $\sigma_w/\sigma_u$ plotted as a function of $d_{\rm max}$.  The
ratio settles to something close to the energy equipartition
prediction when none but the closest heavy scatterers perturb the
stars, but as I increase the range of scattering in separate
experiments, the final ellipsoid gradually becomes flatter and
approaches Ida's result for no cut off.

This test therefore clearly supports the idea that the anisotropic
distribution of the scatterers determines the equilibrium shape of the
velocity ellipsoid.  Previous studies (Spitzer \& Schwarschild 1953;
Lacey 1984; Binney \& Lacey 1988) assumed that scattering is dominated
by impact parameters that are small compared to both the epicycle
radius and the disk thickness and peculiar velocities were the most
important.

Thus the shape of the local velocity ellipsoid (\eg\ Nordstr\"om
\etal\ 2004) is apparently consistent with cloud scattering, and its
origin does not {\it require\/} concurrent spiral arm scattering, as
seemed attractive.  However, the data do not imply that spirals are
unimportant: there are hints of some evolution of the velocity
ellipsoid shape that may demand a compound origin, and the magnitude
of the random speeds of the oldest stars may still require an
additional source of scattering (\eg\ Lacey 1991).

\acknowledgments I thank Scott Tremaine for insisting that I should
better understand the reason for the shape of the local velocity
ellipsoid, and for some helpful correspondence.  I thank Cedric Lacey,
Ray Carlberg, Victor Debattista and especially James Binney for
perceptive comments on the manuscript.  This work was supported by
grants AST-0507323 from the NSF and NNG05GC29G from NASA.

\end{document}